\documentclass[aip,apl,amsmath,amssymb,preprint,groupedaddress]{revtex4-1} 

\usepackage{graphicx}

\begin{document}

%\preprint{AIP/123-QED}
\title[]{Exciton bound by distant ionized donor in two-dimensional GaAs/AlGaAs quantum well} 
\author{M.~Tytus}
 \email{marcin.tytus@pwr.wroc.pl.}
\author{W.~Donderowicz}
\author{L.~Jacak}
\affiliation{Institute~of~Physics, Wroc\l{}aw~University~of~Technology, Wybrze\.ze~Wyspia\'nskiego~27, 50-370~Wroc\l{}aw, Poland}
\date{\today}

\begin{abstract}
The ground state energy of exciton bound by distant ionized donor impurity in quasi-two-dimensional GaAs/Al$_x$Ga$_{1-x}$As semiconductor quantum well (QW) is studied theoretically within the Hartree approach in the effective mass approximation.
The influence of the distance between QW plane and ionized donor, as well as of the magnetic field aligned across the QW plane and varying dielectric constant of the barrier material on the stability of exciton bound by ionized donor impurity
is analyzed and discussed.
\end{abstract}

\pacs{78.67.De, 78.66.Fd, 78.20.Ls, 73.21.Fg}
%\keywords{ionized donor bound exciton, magnetoexciton}
\maketitle

One of the simplest possible three-particle bound-exciton complex $(D^+,X)$ 
consist of an exciton $X$ (electron hole pair) 
bound to an ionized donor impurity $D^+$.
Its possible existence was first predicted by Lampert~\cite{Lampert} in 1958.
Since then its stability and binding energy
has been the subject of few theoretical studies 
in bulk (3D) semiconductor \cite{Skettrup, Stauffer} and 
in two-dimensional QW structures \cite{Stauffer,Ruan}.
In 2D case the overlappings between the wave functions 
of the constituents of the $(D^+,X)$ 
complex 
become more important  (due to the quantum confinement), 
binding energy of the complex is increased
and so the stability in 2D structures is increased 
compared to their 3D counterparts.
Therefore it is expected that the observation of bound excitons 
should be easier in 2D structures than in the bulk.

The binding energy of an exciton 
bound to an ionized donor
in GaAs/Al$_x$Ga$_{1-x}$As QW 
has been calculated for finite well widths
and different impurity position 
by 
Liu and co-workers \cite{Liu3} and 
da Cunha Lima \textit{et al.} \cite{Lima}.
St\'eb\'e and co-workers \cite{Stebe} studied variationally  
the influence of the magnetic field 
on the stability of created complex.
Nevertheless no one has yet analyzed the impact of 
ionized donor shifted from the QW plane.

The lateral crossection of singular potentian of ionized donor, 
as acting on charge carriers in distant 2D well, 
resembles a nonsingular potential of type-II quantum dot (defined by the electrostatic field) 
--- thus recognition of exciton evolution with respect to the donor separation 
is of crucial importance in order to differentiate both confinements. 

In this letter the ground state energy of exciton 
bound by ionized donor shifted from 2D well
is studied theoretically within the Hartree approach 
in the effective mass approximation.
The influence of 
the donor distance,
varying dielectric constant of barrier material
and of the external uniform magnetic field
on the stability 
of 2D exciton bound by ionized donor
is analyzed and discussed.

For the model analysis, we assume 
that the QW is quasi-two-dimensional 
and lies in the x-y plane,
while
the magnetic field
is aligned across this plane, i.e., along the z axis.
We restrict our model 
only to the spatial coordinates
--- spin degrees of freedom 
and the associated Zeeman splitting (linear in B)
were not included in our description
(for GaAs this splitting is very small $\sim 0.03\,$meV/T.)

In the QW plane 
potential of ionized donor shifted by the distance $d$ 
in the axial direction has the form
\begin{equation}
  \label{Vd}
    V_i\left(\rho_i\right) = \mp \frac{q^2}{4\pi\epsilon_1\epsilon_0}\frac{1}{\sqrt{\rho_i^2+d^2}},
\end{equation}
where minus sign corresponds to the electron ($i=e$), plus to the hole ($i=h$),
$\rho_e$ and $\rho_h$ are the radial distances of electron and hole in the plane,
$q$ is the elementary positive charge
and $\epsilon_1$ is the relative dielectric constant of the barrier material.

Despite the fact that the potential~\eqref{Vd}
is attractive only for one type of charge carrier
it is however possible for the distant donor to captured the electron-hole pair (exciton) 
due to Coulomb interaction between charge carriers.

It should be noted here, 
that in more realistic model, for small $d$ and shallow QWs,  
electron tunneling through the potential barrier has to be taken into account.
Nevertheless, the probability of this process rapidly decreases 
as the donor is shifted away from QW plane.

Within the Hartree method exact exciton wave function can be approximated by
$\Psi\left(\textbf{r}_e,\textbf{r}_h\right) = \psi_e\left(\textbf{r}_e\right) \psi_h\left(\textbf{r}_h\right)$.
Using axial symmetry we assume
\begin{equation*}
     \psi_s\left(\textbf{r}_s\right)=\frac{1}{\sqrt{2\pi}}\exp\left(il_s\varphi_s\right)\phi_s\left(\rho_s\right),
\end{equation*}
where $l_s=0,\pm 1,\pm 2,\ldots$ and $s=e,h$.
Then the single-particle Hartree energies and wave functions are found 
in the effective mass approximation
by iterative solving of self-consistent Hartree equations
\begin{align}
\begin{split}
\label{RS}
		\Bigg[&
	  -\frac{\hbar^2}{2m_i}\frac{1}{\rho_i}\frac{\partial}{\partial \rho_i}\left(\rho_i \frac{\partial}{\partial \rho_i}\right) +\frac{\hbar^2}{2m_i}\frac{l_i^2}{\rho_i^2} \\ 
		&+U_i\left(\rho_i\right) +\frac{1}{8}m_i\omega_{ci}^2\rho_i^2 \pm\frac{l_i}{2}\hbar\omega_{ci} 
		\Bigg] \phi_i\left(\rho_i\right) = \varepsilon_i \phi_i\left(\rho_i\right)
\end{split}
\end{align}
with the Hartree patentials
\begin{align}
\label{U}
   U_i\left(\rho_i\right)  =  V_i\left(\rho_i\right)  
    -\frac{q^2}{4\pi\epsilon_2\epsilon_0} \int \frac{\left|\psi_j\left(\textbf{r}_j\right)\right|^2}{\left|\textbf{r}_i-\textbf{r}_j\right|}d\textbf{r}_j
\end{align}
where $i=e$ and $j=h$ (or opposite),
the upper sign corresponds to the electron and the lower sign to the hole,
$m_e$ and $m_h$ are effective electron and hole masses respectively,
$\omega_{ce}=qB/m_e$ and $\omega_{ch}=qB/m_h$ are electron and hole cyclotron frequencies
and $\epsilon_2$ is the relative dielectric constant of the QW material.

The exciton energy in Hartree approximation is given by
$E = \varepsilon_e + \varepsilon_h - V_C$,
where
\begin{equation}
\label{Vc}
		V_C  =  -\frac{q^2}{4\pi\epsilon_2} \iint \frac{\left|\psi_e\left(\textbf{r}_e\right)\right|^2\left|\psi_h\left(\textbf{r}_h\right)\right|^2}{\left|\textbf{r}_e-\textbf{r}_h\right|}d\textbf{r}_e d\textbf{r}_h.
\end{equation}
As we deal with single electron-hole pair 
there is no exchange energy term 
(related to Pauli exclusion principle)
and only correlation energy is omitted.
Moreover,
as it was shown for quantum dots 
(whose potential is similar to the potential of shifted donor in QW plane),
the contribution of the correlation 
to the total energy for single electron-hole pair 
is expected to be less than 2\% \cite{Brasken}.

Hartree equations \eqref{RS} 
were solved numerically with finite difference scheme on nonuniform grid
(more details about the implementation of this finite difference scheme 
can be found in work of Peeters \textit{et al.} \cite{Peeters}).
Using this scheme we obtained symmetric tridiagonal matrix.
Its eigen values were calculated with Martin-Dean algorithm \cite{Dean},
whereas eigen vectors were found using DWSZ method \cite{Dy}.
Hartree integrals in \eqref{U} and \eqref{Vc}
were calculated with use of logarithmically weighted method 
after Janssens \textit{et al.} \cite{Janssens}.

\begin{figure}[tb]
	\centering
	\includegraphics{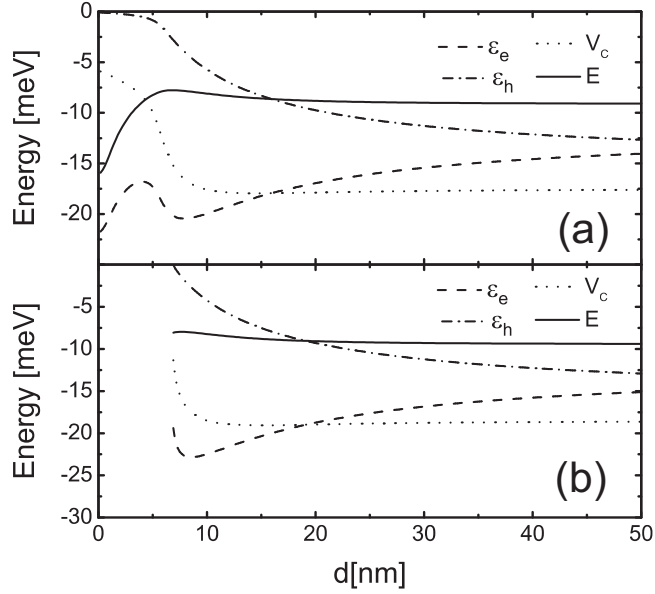}
	\caption{\label{fig:E_d}Electron (dashed line) and hole (dash-dot) Hartree energies,
													coulomb (dot) and
													exciton (solid) energies
	                        as a function of the donor distance $d$ from the QW plane
	                        for $\epsilon_1=\epsilon_2=12.4$ (a)
	                        and $\epsilon_1=10.1$, $\epsilon_2=12.4$ (b).
	        }
\end{figure}

For GaAs semiconductor QW we choose as material parameters
$\epsilon_2=12.4$, $m_e=0.0665$, $m_h=0.3774$.
Fig. \ref{fig:E_d}(a) shows
electron (dashed line) and hole (dash-dot) Hartree energies
as well as
coulomb (dot) and exciton (solid) energies
as a function of the donor distance $d$ from the QW plane for $\epsilon_1=\epsilon_2=12.4$.
As is clearly seen 
exciton energy almost stops changing
for distance values greater than 20$\,$nm, 
for which it should correspond to the energy of free 2D exciton.
For decreasing $d$ the Hartree hole energy is getting smaller,
but as can be shown by extrapolating the energy dependence 
it never reaches zero.

The different situation is for $\epsilon_1<\epsilon_2$.
As we can see in Fig. \ref{fig:E_d}(b) ($\epsilon_1=10.1$ as for AlAs), 
the hole is not bound
until the distance of donor
reaches the critical value $d_{min}$.
It turns out that
the lower dielectric constant of the barrier material,
prevents the exciton binding for small $d$
--- hole is more strongly repelled
and the electron Coulomb attraction is insufficient 
to bound a hole %($\varepsilon_h>0$)
until we move the donor at an appropriate distance.
The question arises
how far to move the donor from the plane of the well
for a given dielectric constant of the barrier
in order to obtain a bound state.

\begin{figure}[tb]
	\centering
	\includegraphics{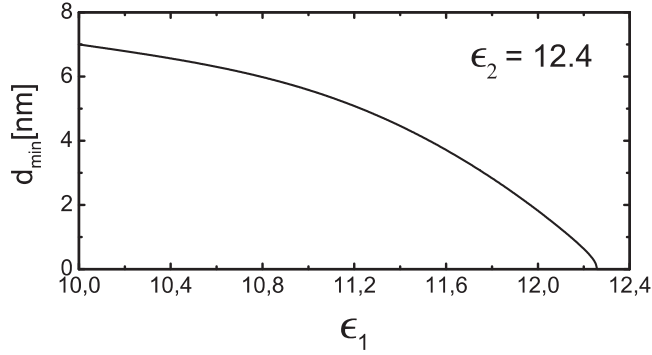}
	\caption{\label{fig:dmin}The dependence of critical distance $d_{min}$ on the $\epsilon_1$.
					}
\end{figure}

Fig. \ref{fig:dmin} shows 
the dependence of this critical distance
on the $\epsilon_1$ for GaAs/Al$_x$Ga$_{1-x}$As,
for which we assume $\epsilon_1(x)=12.4-2.3x$.
As one might expect 
$d_{min}$ 
decreases with increasing $\epsilon_1$
and reaches zero before $\epsilon_1$ equals $\epsilon_2$.

Even if exciton is bound by ionized donor impurity ($\varepsilon_h<0$)
created complex may be unstable
due to the following dissociation processes
\begin{align}
\label{disD}
	(D^+,X) \rightarrow & D^0 + h,\\
\label{disX}
	(D^+,X) \rightarrow & D^+ + X.
\end{align}
In these equations
$(D^+,X)$ and $D^0$ denote respectively exciton or electron bound by ionized donor in the QW,
while
$h$ and $X$ denote free hole and free exciton in the QW plane.
Therefore, we need to consider the binding energies 
$E^B_{D^0} = E_{D^0} + E^f_h - E_{(D^+,X)}$
and
$E^B_X     = E^f_X           - E_{(D^+,X)}$
whose physical meaning is that
the $E^B_{D^0}$ is the minimum energy 
required to liberate the hole from the bound exciton
and $E^B_X$ is the minimum energy required 
to liberate the exciton from the influence of ionized donor.
So the complex remains stable if $E^B_{D^0}>0$ and $E^B_X>0$.

In our calculations
for $E^f_X$ we take the value of bound exciton energy for very large $d$ (for which it stabilizes),
and we put $E^f_X$ equal $\hbar\omega_{ch}/2$ (lowest landau level in magnetic field).

\begin{figure}[tb]
	\centering
	\includegraphics{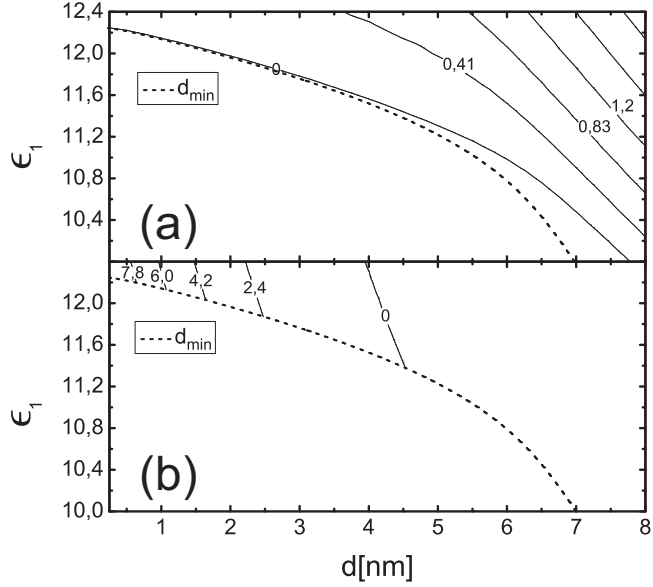}
	\caption{\label{fig:Eb_ed_d}Contour plots of energies $E^B_{D^0}$ (a) and $E^B_X$ (b), 
														  both in meV,			
					                    depending 
												      on the dielectric constant of barrier material $\epsilon_1$ and 
												      the donor distance $d$ from the QW plane.
													    Dotted lines indicate the limit distance for donor.
	        }
\end{figure}

Fig. \ref{fig:Eb_ed_d} represents
the dependence of the energies $E^B_{D^0}$ and $E^B_X$, 
both in meV,
on the donor distance from the QW and
on the dielectric constant of the barrier material
in the range $10<\epsilon_1<12.4$ (as for Al$_x$Ga$_{1-x}$As).
Additional dotted lines indicate 
the limit distance for donor (cf. Fig. \ref{fig:dmin}).
As can be seen by comparing parts (a) and (b) 
the complex is stable when $11.4\leq\epsilon_1\leq 12.4$
and only for $d_{min}<d\lesssim 4\,$nm.
For $d$ greater than $\sim 4\,$nm,
in the whole range of $\epsilon_1$,
complex may dissociate into $X$ and $D^+$
or,
if in addition $d$ is not much larger than the $d_{min}$,
into hole and $D^0$.

\begin{figure}[tb]
	\centering
	\includegraphics{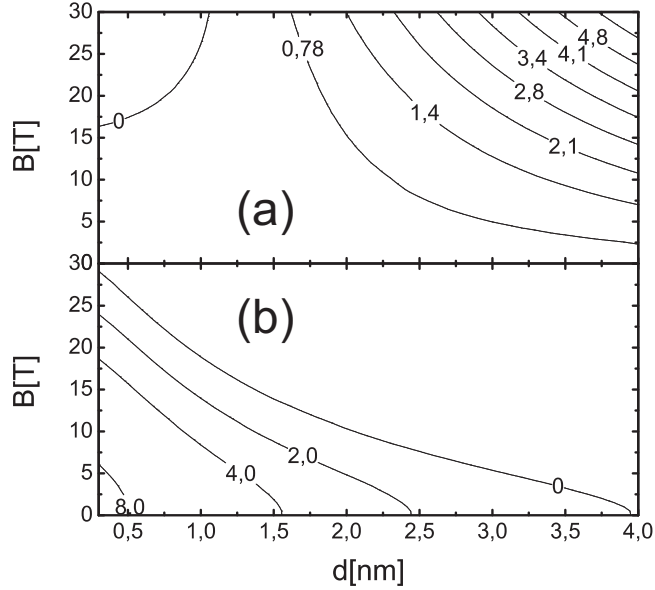}
	\caption{\label{fig:Eb_B_d}The dependence of energies $E^B_{D^0}$ (a) and $E^B_X$ (b),
	                           both in meV, 
														 on the donor distance $d$ from the plane of QW
														 and magnetic field $B$ 
													   for $\epsilon_1=\epsilon_2=12.4$.
	        }
\end{figure}

Fig. \ref{fig:Eb_B_d}
shows the dependence of the binding energies $E^B_{D^0}$ and $E^B_X$
on the donor distance from the QW 
and  magnetic field
for $\epsilon_1=\epsilon_2=12.4$.
As can be seen
$E^B_{D^0}$ is less than zero
only for high magnetic field and respectively small $d$.
In this range of parameters
the complex $(D^+,X)$ is unstable 
due to dissociation process \eqref{disD}.
Moreover,
in the absence of magnetic field
$E^B_X$ is less than zero for $d$ greater than about $3.9\,$nm,
which means that the complex is unstable due to dissociation process \eqref{disX}.
This critical distance decreases with increasing field
which may reflect the fact
that in a magnetic field
the Coulomb interaction energy in $X$ is growing relatively quickly
while the ionized donor (because of repulsive potential for hole)
prevents such rapid growth of this energy in bound exciton.

In summary,
we have found
critical distance at which the donor has to be moved from GaAs/Al$_x$Ga$_{1-x}$As QW plane
in order to bound exciton.
We have also studied stability of created complex
--- it turned out that it is stable
only if $11.4\lesssim\epsilon_1<12.4$,
the distance of the donor $d_{min}<d\lesssim 4\,$nm
and the amplitude of magnetic field is sufficiently small.

\providecommand{\noopsort}[1]{}\providecommand{\singleletter}[1]{#1}%

\end{document}